\documentclass[pdflatex,sn-nature]{sn-jnl}%

\usepackage{graphicx}
\usepackage{multirow}
\usepackage{amsmath,amssymb,amsfonts}
\usepackage{amsthm}
\usepackage{mathrsfs}
\usepackage[title]{appendix}
\usepackage{xcolor}
\usepackage{textcomp}
\usepackage{manyfoot}
\usepackage{booktabs}
\usepackage{algorithm}
\usepackage{algorithmicx}
\usepackage{algpseudocode}
\usepackage{listings}
\usepackage[version=3]{mhchem}
\usepackage{lineno}
\usepackage{anyfontsize}

\begin{document}

\title[Article Title]{Termination of bottom-up interstellar aromatic ring formation at \ce{C6H5+}}

\author*[1,2]{\fnm{G.~S.} \sur{Kocheril}}\email{stephen.kocheril@colorado.edu}

\author[1,2]{\fnm{C.} \sur{Zagorec-Marks}}

\author[1,2]{\fnm{H.~J.} \sur{Lewandowski}}

\affil[1]{\orgdiv{JILA}, \orgname{National Institute of Science and Technology}, \orgaddress{\city{Boulder}, \postcode{80309}, \state{CO}, \country{USA}}}

\affil[2]{\orgdiv{Department of Physics}, \orgname{University of Colorado, Boulder}, \orgaddress{\city{Boulder}, \postcode{80309}, \state{CO}, \country{USA}}}

\abstract{The aromatic molecule benzene is considered the essential building block for larger polycyclic aromatic hydrocarbons (PAHs) in space. Despite benzene's importance in the formation of PAHs, the formation mechanisms of interstellar benzene are not well understood. A single ion-molecule reaction sequence is considered when modeling the formation of benzene in the interstellar medium, beginning with the protonation of acetylene. Although this process has been used to model the initial steps for formation of PAHs, it has not been experimentally measured. To explore this reaction mechanism, we have carried out the first experimental study of sequential ion-molecule reactions beginning with protonation of acetylene at single-collision conditions. Surprisingly, we find that the reaction sequence does not result in benzene and instead terminates at \ce{C6H5+}, which is unreactive toward both acetylene and hydrogen. This result disproves the previously proposed mechanism for interstellar benzene formation, critically altering our understanding of interstellar PAH formation.}


\maketitle
\newpage
\section*{Introduction} 
Aromatic molecules play a critical role in our understanding of the chemical evolution of the universe~\cite{allamandola2021pah}. About 10 -- 25\% of all interstellar carbon is assumed to be in the form of polycyclic aromatic hydrocarbons (PAHs), which are attributed to be the source of numerous broad infrared spectral features found throughout the interstellar medium (ISM)~\cite{tielens2008Review1}. Although PAHs were long inferred to be present in sizable abundance in the ISM, recent observations have definitively proven the existence of a number of aromatic molecules by rotational spectroscopy at radio frequencies~\cite{mcguire2018C6H5CN,mcguire2021C10H7CN,burkhardt2021C9H8,sita2022C9H7CN,cernicharo2021cylic,agundez2023aromatic}. Benzene, the building block for all PAHs, has been observed in a variety of astronomical environments including: planetary nebula~\cite{cernicharo2001benzene}, post-Asymptotic Giant Branch stars~\cite{kraemer2006benzene}, circumstellar envelopes of evolved carbon-rich stars~\cite{malek2011benzene}, and the comae of comets and asteroids~\cite{schuhmann2019benzene}. Benzene has also been observed in the atmospheres of Saturn~\cite{koskinen2016benzene} and Titan~\cite{waite2007benzene}, as well as meteoritic chondrites~\cite{delsemme1975benzene}. Recent observations by the James Webb Space Telescope have also revealed the presence of benzene, as well as other smaller hydrocarbons, in protoplanetary discs~\cite{tabone2023JWST,arabhavi2024JWST}. These observations have reinvigorated interest in interstellar benzene and PAHs because the formation mechanisms of PAHs throughout the ISM remain an open question. Exact mechanisms of PAH formation cannot be understood from astronomical observations alone. Instead, laboratory measurements are needed to supplement observations and chemical models. To this end, we report the first experimental mechanistic study of the sequential ion-molecule reactions starting with protonated acetylene and acetylene under single-collision conditions to identify a possible mechanism to form benzene in the ISM.

\begin{figure}
    \centering
    \includegraphics[width=0.9\linewidth]{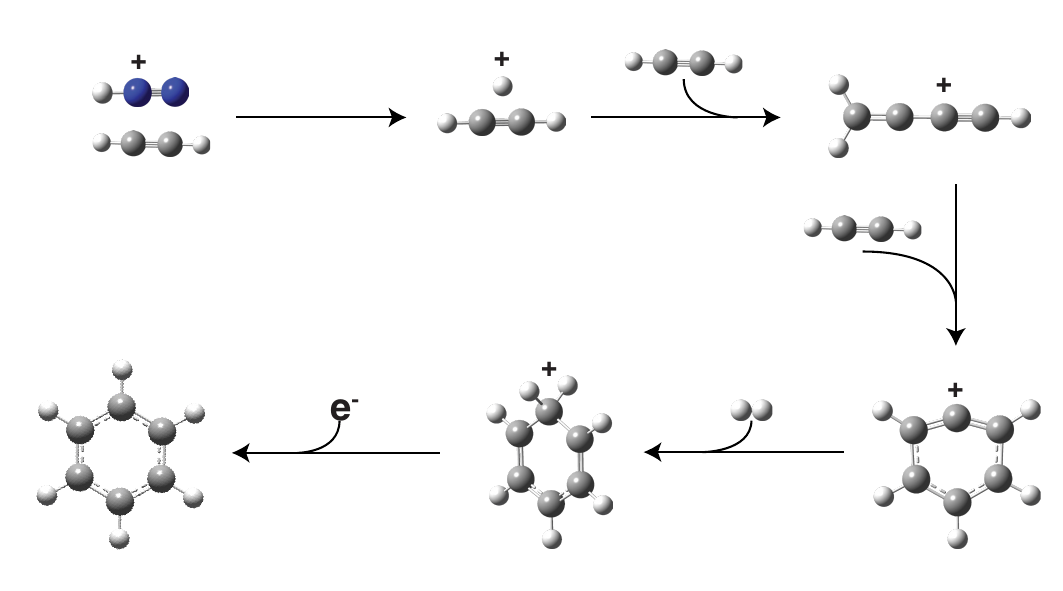}
    \caption{\textbf{Proposed synthetic mechanism to form benzene via ion-molecule reactions.} A proton donor (such as \ce{N2H+}) initiates the mechanism by protonating \ce{C2H2} thereby forming \ce{C2H3+}. \ce{C2H3+} then reacts with \ce{C2H2} to form \ce{C4H3+} via a condensation reaction. \ce{C4H3+} then undergoes a radiative-association reaction with \ce{C2H2} to form \ce{C6H5+}. \ce{C6H5+} then undergoes a radiative-association reaction with \ce{H2} to form \ce{C6H7+}. Finally, \ce{C6H7+} undergoes electron recombination to form benzene.}
    \label{fig:Ring Formation}
\end{figure}

Formation mechanisms of PAHs fall into two main categories: top-down mechanisms where large interstellar clusters or large particles are fragmented by UV radiation to form PAHs~\cite{berne2015topdown}, or bottom-up mechanisms where small hydrocarbon molecules undergo a series of barrierless ion-molecule or radical-radical reactions to grow into large PAHs~\cite{woods2002benzeneformation,jones2011rr,pentsak2024C2H2}. Although there is still an active debate over which mechanism is dominant in the context of the ISM, the bottom-up approach has received more attention due to the recent detection of PAHs specifically in TMC-1 CP, a dark and diffuse starless cloud located within the interstellar Taurus Molecular Cloud, where it is highly unlikely that PAH formation would be dominated by the degradation of large clusters or carbonaceous macromolecules~\cite{mcguire2021C10H7CN,chabot2019PAHdestruction}. 
There have been several experiments over the past decade that have identified multiple possible bottom-up-based PAH growth mechanisms, including ion-molecule reactions, radical-radical reactions, and radical-molecule reactions~\cite{rap2024noncovalent,kaiser2021aromatic,bierbaum2011pahs}. All of these mechanisms begin with an already established aromatic molecule (one or more rings). Surprisingly, the formation of the first aromatic ring has not received much experimental attention~\cite{lee2019C6H5CNformation}. In bottom-up-focused PAH formation mechanisms, the formation of the first aromatic ring (i.e., benzene) is widely considered both the limiting step of the process and the most difficult to probe experimentally~\cite{kaiser2021aromatic,lee2019C6H5CNformation}. Typically, astrochemical models must consider both radical-radical reactions and ion-molecule reactions to accurately reproduce observed molecular abundances. One primarily ion-molecule based bottom-up mechanism has been considered for benzene formation throughout the ISM~\cite{mcewan1999benzeneH2mech,woods2002benzeneformation}, shown schematically in Figure~\ref{fig:Ring Formation}, where the reaction is initiated by the protonation of acetylene (\ce{C2H2}) by a proton donor, e.g. \ce{H3+}, \ce{N2H+}, \ce{HCO+}. The newly formed \ce{C2H3+} then reacts with another \ce{C2H2}, undergoing a condensation reaction to form \ce{C4H3+}, followed by another addition of \ce{C2H2} to undergo a radiative-association-type reaction to form \ce{C6H5+}. It is assumed that the structure of \ce{C6H5+} is that of the deprotonated benzene cation (phenylium), which can undergo another radiative-association reaction with molecular hydrogen (\ce{H2}) to directly form protonated benzene cation (\ce{C6H7+}). Finally, protonated benzene can undergo electron recombination to yield neutral benzene. Although this mechanism has been included in chemical models~\cite{mcewan1999benzeneH2mech,woods2002benzeneformation}, this series of reaction has not been experimentally verified from start to finish. The reactivity of \ce{C6H5+}, specifically, has not been studied at low densities and temperatures, making it difficult to predict its reactivity in the ISM.

There has been prior experimental evidence to support the individual steps of this mechanism. Multiple types of ion-molecule reaction experiments have observed that sequential ion-molecule reactions of \ce{C2H2+} + \ce{C2H2} result in the formation of both \ce{C4H2+} and \ce{C4H3+} as primary products and both \ce{C6H4+} and \ce{C6H5+} as secondary products~\cite{brill1981sequential,myher1968C2H2,futrell1968C2H2,anicich1986C2H2,knight1987SIFT}. Additionally, multiple isomers of \ce{C6H5+} were observed in these experiments, where one isomer was observed to be highly reactive with \ce{C2H2} thereby forming \ce{C8H6+} and \ce{C8H7+}, and another isomer was found to be unreactive. The observation of multiple isomers prompted an assignment of the unreactive isomer to be the phenylium structure and the reactive isomer to be an acyclic structure. The assignment of isomer structures was based on an energetics argument, as the cyclic isomer was known to be far lower in energy than the acyclic structure~\cite{anicich1986C2H2,knight1987SIFT}. 

The formation of a possible phenylium cation and its lack of reactivity was of high interest to chemists generally. This led to several experiments with the express purpose of understanding both the structure and reactivity of the \ce{C6H5+} ion. These results were somewhat controversial, as multiple studies found that the assumed phenylium isomer of \ce{C6H5+} was unreactive~\cite{eyler1984C6H5+,giles1989C6H5+,contreras2013acetylenereact}, while several others found that is was highly reactive with acetylene~\cite{fornarini1985phenylium,ausloos1989c6h5+,soliman2012sequential}, \ce{H2}~\cite{petrie1992c6h5+,scott1997c6h5+,ascenzi2003C6H5}, and other hydrocarbons~\cite{speranza1977phenylium,lifshitz1980c6h5+,ascenzi2007benzene}. Multiple collisionally induced dissociation studies also found that the structure of a \ce{C6H5+} fragment was highly dependent on the structure of the precursor molecule~\cite{schroder1999charge}. Recent studies on the fragmentation of substituted phenyl rings, phenylbromide or benzonitrile, observed \ce{C6H5+} as a major product, confirming the phenylium structure with infrared multi-photon dissociation~\cite{wiersma2021C6H5+ir}, vacuum ultra-violet photodissociation~\cite{jacovella2022ultraviolet}, and infrared pre-dissociation (IRPD)~\cite{rap2023C6H5+IR}. To our knowledge, there have been no spectroscopic studies on the cyclic isomer of \ce{C6H5+} formed via bottom-up synthesis. 

All of the previously mentioned reaction experiments were conducted at high pressures (10$^{-6}$ Torr or higher) and most were also at high collision energies ($>$1 eV). These environments likely led to the variability in the perceived reactivity of the \ce{C6H5+} ion, as at higher pressures, collisional stabilization will lead to a greater abundance of higher energy structures, which are more reactive than the global minimum structure. High collision energies make it easier to break chemical bonds, thereby unknowingly overcoming reaction barriers. Experiments carried out at these pressures and collision energies far exceed what is present in the ISM, where densities are low enough that termolecular interactions effectively never occur. This means that our current understanding of aromatic ring formation is hindered by incomplete experimental data. To further refine models of interstellar chemical reactions, especially bottom-up-focused ion-molecule mechanisms, a new type of experimental system that mirrors the environment of the ISM is required.

\begin{figure}
    \centering
    \includegraphics[width=0.9\linewidth]{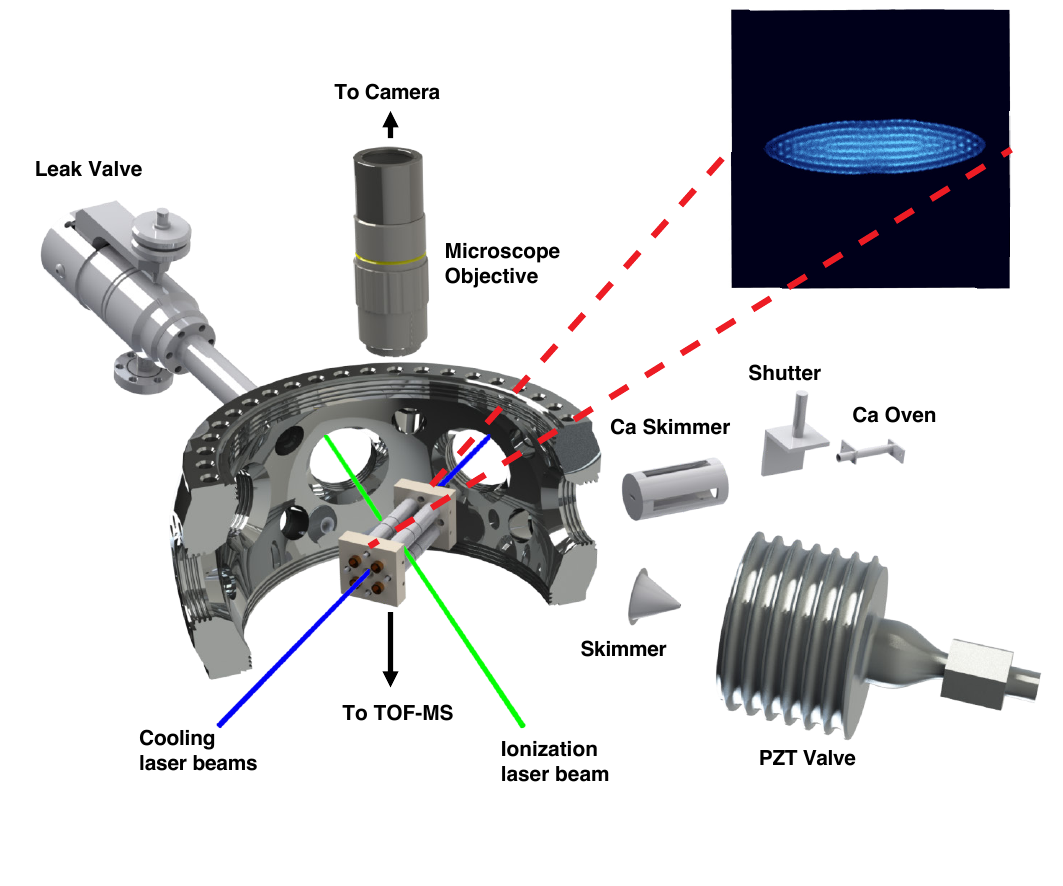}
    \caption{\textbf{Schematic of the experimental apparatus}. A cross-sectional view of the UHV chamber is shown with the main experimental components. An effusive Ca atomic beam is non-resonantly photoionized and the resulting ions are confined within a linear Paul trap. Molecular ions are loaded by resonantly photoionizing a skimmed molecular beam created by a PZT valve. Fluorescence from laser-cooled \ce{Ca+} ions are imaged with an EM-CCD camera (an example of such an image with co-trapped \ce{C2H3+} ions is shown in the inset) providing a qualitative observation of the reaction's progression. Neutral reactants are admitted through the leak valve at a fixed pressure for variable amounts of time before the contents of the trap are ejected into the time-of-flight mass spectrometer and detected using a set of microchannel plates in a chevron configuration. A new Coulomb crystal is loaded for each measurement. Adapted from Schmid et al.\cite{schmid2020isomer}. Copyright 2020 PCCP Owner Societies.}
    \label{fig:Apparatus}
\end{figure}

The experimental system we use is able to mimic the low pressure and low energy environment of the ISM by taking advantage of the techniques of trapping and laser-cooling of atomic ions in an ultra-high vacuum chamber, shown in Figure~\ref{fig:Apparatus}. To reach single collision conditions and temperatures below 10 K, we use a linear quadrupolar Paul trap to confine and laser-cool \ce{Ca+} in a Coulomb crystal, which is a pseudo-crystalline structure that is formed by an ensemble of cold, trapped ions, shown in the inset of Figure~\ref{fig:Apparatus}. The cold \ce{Ca+} ions sympathetically cool co-trapped molecular ions to single Kelvin translational temperatures~\cite{schmid2017apparatus}. As the Coulomb crystal has $\sim 10 \mu m$ ion-ion spacing, there is no collisional cooling of vibrational or rotational modes for the co-trapped molecular ions. A detailed description of the apparatus is provided in the Methods. Briefly, reactions are initiated by introducing the neutral reactant into the vacuum system at pressures that result in collision rates between the neutral acetylene and trapped ions on the order of one collision per second. This ensures that we have only two-body collisions and do not need to consider the possibility of termolecular or greater reactions. As a reaction proceeds, all ionic products are retained in the trap and we analyze the contents of the trap by ejecting all ions into a time-of-flight mass spectrometer. By measuring the number of ions with each mass in the trap at different times, we are able to build kinetic reaction curves and determine the rates for each sequential reaction using a pseudo-first order rate law, as the neutral \ce{C2H2} is in excess. The low densities and temperatures of both the ion and neutral reactants allows us to simulate the conditions similar to that of the ISM and thus study both mechanisms and rates relevant to that environment. We find that the series of sequential reactions beginning with \ce{C2H3+} +\ce{C2H2} does not result in benzene and instead stops at \ce{C6H5+}, which is unreactive for reactions with both \ce{C2H2} and \ce{H2}. These results demonstrate the established bottom-up mechanism for benzene formation does not actually produce benzene at single-collision conditions.

\section*{Results}
The series of ion-molecule reactions are initiated by the protonation of \ce{C2H2}. The protonation of hydrocarbons by \ce{N2H+} and other proton donors has been studied in detail previously~\cite{milligan2002PTR}, and it is well understood that protonation of \ce{C2H2} by \ce{N2H+} or even stronger acids like \ce{H3+} results in the production of only \ce{C2H3+}~\cite{milligan2002PTR,kim1974H3+}. 

\begin{figure} 
	\centering
	\includegraphics[width=0.9\linewidth]{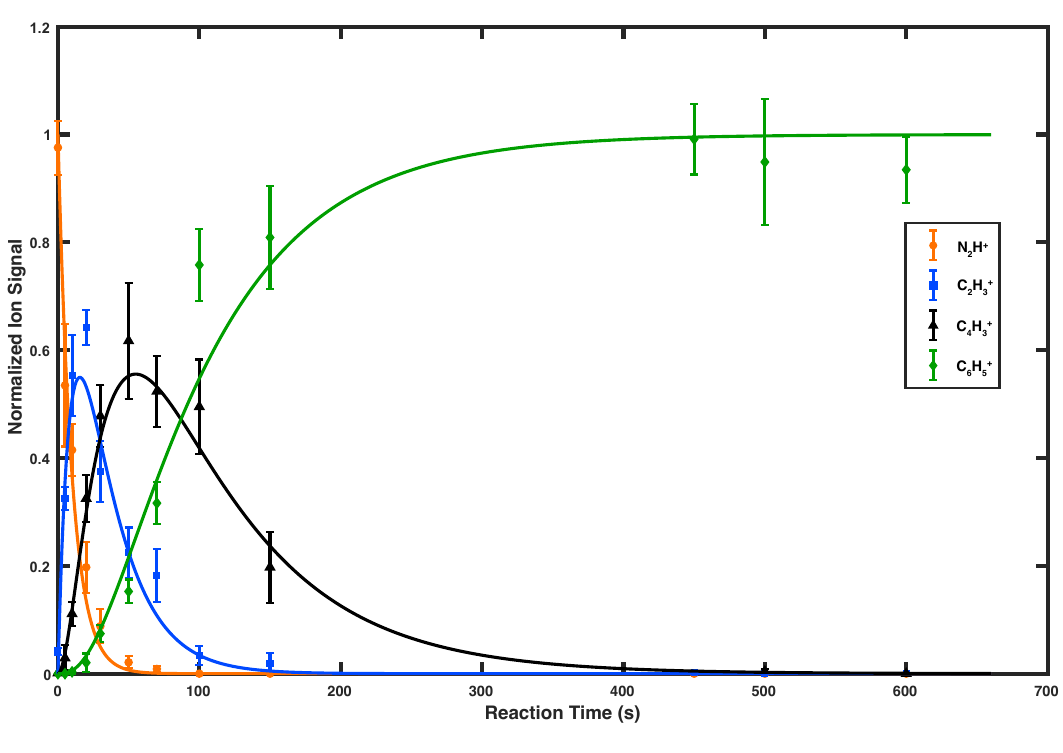} 
	\caption{\textbf{Ion reactant depletion and product growth for the sequential chain reactions starting with \ce{N2H+} + \ce{C2H2}.} Experimental data (points) are fit (curves) using a pseudo-first order rate law. Each data point corresponds to seven measurements and is normalized to the initial number of \ce{N2H+} ions ($\sim90$). The orange circles corresponding to \ce{N2H+} exhibit a sharp decay indicating a rapid reaction of the ions with \ce{C2H2}, producing \ce{C2H3+}. As \ce{C2H3+} (blue squares) is produced, it reacts with another \ce{C2H2} to form \ce{C4H3+}. The final reaction observed is \ce{C4H3+} (black triangles) reacting with another \ce{C2H2} to produce \ce{C6H5+}. The \ce{C6H5+} (green diamonds) signal is observed to plateau over time indicating the termination of the chain reactions. All error bars represent the 68\% confidence interval from the fit.}
	\label{fig:Reaction Curves} 
\end{figure}

The product growth observed for the chain reactions are shown in Figure~\ref{fig:Reaction Curves}. The series of observed reactions are listed in Equations 1 -- 3.

\begin{equation}
	\ce{N2H+} + \ce{C2H2} \xrightarrow{} \ce{C2H3+} + \ce{N2}
	\label{eq:Reaction1} 
\end{equation}

\begin{equation}
	\ce{C2H3+} + \ce{C2H2} \xrightarrow{} \ce{C4H3+} + \ce{H2}
	\label{eq:Reaction2} 
\end{equation}

\begin{equation}
	\ce{C4H3+} + \ce{C2H2} \xrightarrow{} \ce{C6H5+} + h\nu
	\label{eq:Reaction3}
\end{equation}

\noindent 
\ce{N2H+} (\textit{m/z}=29) reacts with neutral \ce{C2H2} and results in a reduction of \ce{N2H+} ions. The protonation of \ce{C2H2} leads to the formation of \ce{C2H3+}(Eq.~\ref{eq:Reaction1}), indicated by the rise of \ce{C2H3+} (\textit{m/z}=27) signal at early times. \ce{C2H3+} then reacts with \ce{C2H2} to form \ce{C4H3+}(Eq.~\ref{eq:Reaction2}), shown by the rise of the \ce{C4H3+} (\textit{m/z}=51) ion signal and decay of \ce{C2H3+}. The final observed reaction is the formation of \ce{C6H5+} via reactions between \ce{C4H3+} and \ce{C2H2} (Eq.~\ref{eq:Reaction3}), shown by the slow increase and plateau of \ce{C6H5+} (\textit{m/z}=77) ion signal. The rates of each reaction are determined from the fits of each reaction set. The rate constants for each reaction are provided in Supplementary Table 1, along with comparisons to literature values and Langevin theory. 

\begin{figure} 
	\centering
	\includegraphics[width=0.9\textwidth]{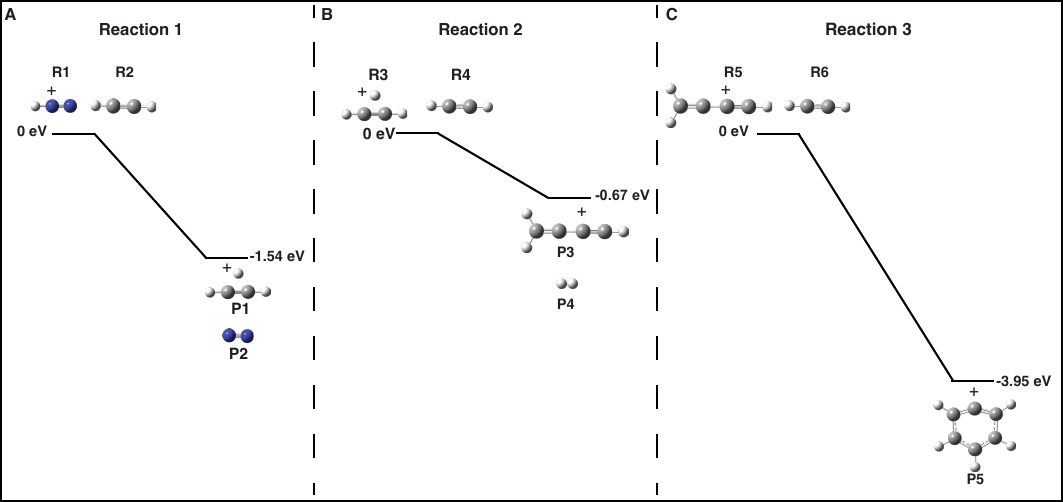} 

	\caption{\textbf{A schematic diagram of the full series of sequential reactions.} Only reactant and product states are shown for clarity. Reaction 1 (A) shows protonation of acetylene (R2) by \ce{N2H+}(R1), which is calculated to be 1.54 eV exothermic. In Reaction 2 (B), \ce{C2H3+} (R3) reacts with \ce{C2H2} (R4) and undergoes a condensation reaction to form \ce{C4H3+} (P3), which is calculated to be 0.67 eV exothermic. The final reaction, Reaction 3 (C), \ce{C4H3+} (R5) reacts with \ce{C2H2} (R6) to form the final observed product \ce{C6H5+} (P5) through a radiative-association reaction, which is calculated to be 3.95 eV exothermic. Each reaction is computed at MP2/aug-cc-pVTZ level of theory. The full potential energy surface for these reactions are shown in Supplementary Figures 1 and 2.}
	\label{fig:Sequential Rxns} 
\end{figure}

Although we are unable to probe the structures of the ions experimentally, we are able to utilize quantum chemical calculations and prior spectroscopic work on these ions to infer the structures of the ions produced throughout the reactions and construct a potential energy surface (PES) (see Supplementary Figures 1 and 2 for details). Beginning with the primary product, \ce{C2H3+}, the structure has been probed in detail through spectroscopy in the infrared~\cite{crofton1989ir,douberly2008ir}, as well as millimeter and sub-millimeter wavelengths~\cite{bogey1992c2h3mmsmm}. The structure is universally found to be the ``nonclassical" or bridged structure. Calculations also support the conclusion that this structure is the only energetically accessible structure in our experiments, so we assign the \ce{C2H3+} structure to be the bridged structure. This reaction is shown in panel A of Figure~\ref{fig:Sequential Rxns}. 

The secondary product, \ce{C4H3+}, labeled as P3 in Figure~\ref{fig:Sequential Rxns}, has been studied in far less detail. Calculations have found that there are five possible structures for \ce{C4H3+}~\cite{peverati2016PES}, but only the linear, protonated diacetylene structure is energetically accessible in our experiments, which is $-0.68$ eV exothermic. The other computed structures were either endothermic by at least $0.5$ eV or greater, or found to have sizable reaction barriers~\cite{peverati2016PES}. There have been two spectroscopic studies of \ce{C4H3+} isomers; one was carried out in He nanodroplets, finding that only the linear structure was observed~\cite{moon2023c4h3}. The other was an IRPD study~\cite{rap2023C6H5+IR} where the linear structure of \ce{C4H3+} was observed as a dissociation fragment of benzonitrile. Considering both calculations and the spectroscopic data, we assign the structure of \ce{C4H3+} as the protonated diacetylene structure. 

The final tertiary product is \ce{C6H5+}, labeled P5 in Figure~\ref{fig:Sequential Rxns}. The \ce{C6H5+} cation has been the subject of several theoretical studies~\cite{dill1977C6H5theory,shi2016isomerization,peverati2016PES}. We have carried out additional calculations on possible isomers for our experiments. Over 30 possible isomers are energetically accessible in our experiment~\cite{shi2016isomerization,peverati2016PES}. All spectroscopic data on \ce{C6H5+} universally finds the structure to be the phenylium structure, which is also the global minimum~\cite{wiersma2021C6H5+ir,jacovella2022ultraviolet,rap2023C6H5+IR}. These spectroscopic studies were carried out by fragmenting substituted benzenes, such as benzonitrile, so it difficult to conclusively say that the structure observed in these experiments would be the same structure produced in our experiments due to the difference in formation mechanism. However, we have experimental evidence that strongly suggests the presence of only the phenylium structure. 

Radiative-association products, such as \ce{C6H5+}, are rare in single-collision experiments because most ions formed through highly exothermic reactions will readily break apart into smaller fragments in the absence of stabilizing collisions. In fact, even in the ion-molecule reaction studies that first observed \ce{C6H5+}, it was assumed that the ion was a collisionally stabilized complex~\cite{myher1968C2H2,futrell1968C2H2,brill1981sequential,anicich1986C2H2,knight1987SIFT}. However, our experiments are run under single-collision conditions, meaning collisionally stabilized products are not possible. This means that the structure of the \ce{C6H5+} cation must be able to quench any internal energy via radiative relaxation, as the Coulomb crystal is only able to quench translational energy. Of all computed structures, the phenylium structure seems most reasonable to be able to survive in the trap,  as aromatic rings are known to be highly stable and resistant to fragmentation, as they have many vibrational modes that can dissipate energy through radiative relaxation instead of undergoing unimolecular decomposition~\cite{peverati2016PES}. Additionally, calculations have shown that isomerization of \ce{C6H5+} from various structures to the phenylium structure are energetically feasible in our experiments~\cite{shi2016isomerization,peverati2016PES}. As there are no collisions to stop the isomerization process from happening, it is highly likely that once any isomer of \ce{C6H5+} is formed, it will quickly isomerize into the phenylium structure and then dissipate any excess internal energy through vibrational radiative relaxation before collisions with a neutral partner can occur.

Based on our results, astrochemical models are still missing crucial information to predict the formation of interstellar PAHs. Although our results closely follow the reaction mechanism shown in Figure~\ref{fig:Ring Formation} up to the formation of \ce{C6H5+}, the lack of reactivity with acetylene does not address whether or not it is possible to form benzene using other neutral molecules. To further investigate the later steps of the proposed bottom-up mechanism, we carried out additional experiments where \ce{C6H5+} was formed as described previously, but then instead of continuing to introduce acetylene, we introduced \ce{H2} into the trap to measure the reactivity of \ce{C6H5+} with \ce{H2}. Surprisingly, neither a depletion of \ce{C6H5+} nor appearance of new mass channels were observed, indicating that \ce{C6H5+} did not react with \ce{H2}. The results of this reaction are shown in Supplementary Figure 3. Our measurements put an upper bound on the reaction rate constant for \ce{C6H5+} with \ce{H2} to be $\sim 1\times10^{-12}$ $\frac{cm^{3}}{s}$. This upper limit is one order of magnitude higher than what has been measured previously and three orders of magnitude longer than the rate estimated from Langevin theory~\cite{ascenzi2003C6H5}. The reaction rate constants are summarized in Supplementary Table 1. 

\section*{Discussion and conclusions}
Most of the ions present in this study have not been detected in the ISM, with only \ce{C2H3+} being detected~\cite{muller2024c2h3astro}, despite multiple observations of both acetylene and benzene. Our results strongly suggest that each ion, \ce{C2H3+}, \ce{C4H3+}, and \ce{C6H5+} should be present throughout the ISM, particularly in regions where strong proton donors are present. Therefore, all ions observed in this study should be high interest targets for high-resolution spectroscopy, either in the infrared or the microwave, to enable astronomical observations of these ions. Observation or lack thereof for these ions in various regions of space will give crucial insight for reactions occurring throughout the ISM, as \ce{C6H5+} could be a precursor to the formation of astronomically observed aromatic molecules, such as benzonitrile or indene. Observation of any or all of the product ions from our study would aid in improving astrochemical models, which in turn would improve our understanding of the chemical reactions occurring throughout space. 

The observation of no reaction between \ce{C6H5+} and \ce{H2} is an important finding generally. Initial descriptions of the reactivity of the phenylium cation assumed it would be highly reactive due to the strong electrophilicity of the lowest unoccupied molecular orbital, a vacant nonbonding $\sigma$ orbital, labeled as LUMO in Supplementary Figure 4, which induces sp hybridization of the lone C-atom~\cite{dill1977C6H5theory,wiersma2021C6H5+ir}. On the other hand, the electronic structure of \ce{C6H5+} does support it being highly stable and resistant to chemical reactions. The ground state is described as a closed-shell $^1$A$_1$, as well as aromatic following H{\"u}ckel's $4n+2$ electron counting rule. In fact, the exact same aromatic $\pi$--bonding orbitals of benzene are found in the phenyl cation, outlined in Supplementary Figure 4. As is the case with benzene, it would be expected that phenylium would be highly stable and would react only with molecules that are able to activate the C-C bond, such as neutral radicals or anions, e.g. strong nucleophiles. Prior reaction studies were unable to ascertain which property, electrophilicity or aromaticity, would be the driving force for the reactivity of phenylium, as reactivity was observed to vary depending on experimental conditions~\cite{eyler1984C6H5+,giles1989C6H5+,contreras2013acetylenereact,fornarini1985phenylium,ausloos1989c6h5+,soliman2012sequential,petrie1992c6h5+,scott1997c6h5+,ascenzi2003C6H5,speranza1977phenylium,lifshitz1980c6h5+,ascenzi2007benzene}. However, our results clearly show that the stability enhanced by aromaticity is the dominant factor for the reactivity of phenylium, as no reaction occurs. This means the formation mechanism shown in Figure~\ref{fig:Ring Formation} is impossible at ISM conditions.

It should be noted that the initial formation mechanism shown in Figure~\ref{fig:Ring Formation} is correct up to the formation of \ce{C6H5+}. The sequence of reactions proceed in a barrierless manner and the estimated rates of each reaction are in good agreement with our results. Additionally, calculations suggested that the reaction; 

\begin{equation}
    \ce{C6H5+} + \ce{H2} \xrightarrow{} \ce{C6H7+} +h\nu
\end{equation}

is highly exothermic ($>$2 eV), which would complete the synthetic route to benzene after electron recombination~\cite{mcewan1999benzeneH2mech,woods2002benzeneformation}. Despite this, we observed that \ce{C6H5+} does not react with either acetylene or molecular hydrogen. This implies there are molecular dynamics determining the reactivity and not simply energetics, as there is no barrier to reaction. The reactions of \ce{C6H5+} and other molecules should be of high interest to astrochemists broadly, especially dynamicists, as the exact details of the dynamics on the potential energy surface will likely hold the key to understanding this reaction and improve understanding of chemical reactions between aromatic species, both terrestrial and interstellar. At present, it is clear that the formation of benzene does not occur through sequential ion-molecule reactions with acetylene, as the terminal product \ce{C6H5+} is a highly stable aromatic molecule that is unreactive toward both acetylene and hydrogen. Further studies are required to determine if other bottom-up reactions might form benzene- and by extension PAHs. 

\section*{Methods}
\subsection*{Experimental Methods}
Reactions were conducted in the ion trap apparatus previously described in \cite{schmid2017apparatus}. The core of the apparatus is a linear quadrupolar Paul trap. An effusive beam of atomic Ca is produced from an oven source and directed at the center of the ion trap, where it is non-resonantly ionized ($\sim$7 $\frac{mJ}{pulse}$, 355 nm, 10 Hz) to form \ce{Ca+}. As ions are formed in the trap, they are laser cooled to sub-Kelvin temperatures ($\sim$100 mK) using the doubled output of a Ti-Sapph laser (MSquared SolsTiS, $\sim$397 nm, 2 mW) and an 866 nm diode laser (Toptica DLC Pro, 2 mW). \ce{Ca+} ions are used to sympathetically cool molecular ions to translational temperatures ranging from sub-Kelvin to $\sim20$ K depending on the mass of the molecular ion. The ultra-high vacuum chamber housing the ion trap has a background pressure of $\sim3\times10^{-10}$ Torr, as measured by a Bayard-Alpert ionization gauge. The total number of ions in the trap was on average 1400 \ce{Ca+} ions and $\sim90$ \ce{N2H+} ions. 

To generate the \ce{N2H+} ions used to initiate the series of reactions, \ce{N2+} ions are first loaded into the trap via a $2+1$ resonance-enhanced multiphoton ionization scheme~\cite{kompa1990} ($\sim$237 nm, 0.7 $\frac{mJ}{pulse}$, 10 Hz). Once the \ce{N2+} ions are loaded into the trap, they are allowed  react with trace residual water in the vacuum chamber over the course of two minutes to form the \ce{N2H+} used in this study. After this two minute period had passed, neutral acetylene (99.6\% acetylene dissolved in acetone, Matheson Gas) mixture (20\% in He) is introduced into the chamber through a pulsed leak valve for various reaction times. As the neutral gas is introduced at room temperature, the collision energies between the trapped ions and neutral gas are estimated to be between 150 -- 230 K in the center-of-mass frame. When neutral gas is introduced, the total pressure was measured to be $7.0\pm0.5\times10^{-9}$ Torr with a calculated partial pressure for \ce{C2H2} of $1.4\pm0.1\times10^{-9}$ Torr. After a predetermined reaction time passes, the valve is closed to end the reaction. The contents of the trap are sympathetically cooled for an additional two minutes before being ejected into the time-of-flight mass spectrometer (TOF-MS) for product identification. All ions are considered vibrationally cold, as they have a full second to radiatively decay to the vibrational ground state. The rotational state distribution is assumed to be room temperature. Reactions curves are obtained by repeating this process for various reaction times, ranging from 0 seconds to 10 minutes. Each time point is sampled at least seven times.

For the \ce{C6H5+} + \ce{H2} reaction, \ce{C6H5+} is produced using the same procedure, using a reaction time of 280 seconds. After the two minute sympathetic cooling period, a neutral \ce{H2} (99.99\% Air Liquide) mixture (10\% in He) is admitted into the chamber at at total pressure of $6.5\times10^{-8}$ ($6.3\times10^{-9}$ partial pressure of \ce{H2}). The reaction times are sampled between 0 and 300 seconds before ejecting the contents of the trap into the TOF-MS for product identification. This procedure is repeated at least four times per reaction time.

\subsection*{Theoretical Methods}
\subsubsection*{Rate Equations}
Reaction curve data were fit using a least-squares curve fit and the following pseudo-first-order rate equations:
\begin{equation}
\frac{dN_{29}}{dt} = -k_{1}*N_{29}
\end{equation}
\begin{equation}
\frac{dN_{27}}{dt} = k_{1}*N_{29}-k_{2}*N_{27}
\end{equation}
\begin{equation}
\frac{dN_{51}}{dt} = k_{2}*N_{27}-(k_{3}+k_{4})*N_{51}
\end{equation}
\begin{equation}
\frac{dN_{77}}{dt} = k_{3}*N_{51}
\end{equation}

\noindent
where N$_{i}$ corresponds to the number of ions observed at mass-to-charge ratio i, and k$_{i}$ corresponds to the fitted value of the i-th reaction rate constant of the sequential reactions. The reaction rate constant, k$_{4}$, represents a contaminant protonation reaction occurring between the \ce{C4H3+} ions and trace amounts of acetone present in our acetylene cylinder. The only observed product involving acetone is protonated acetone (\ce{C3H7O+}). This product is formed after the production of \ce{C4H3+} and continues to grow until \ce{C4H3+} ions have been depleted, where they remain in the trap for the duration of the reaction. As the total number of ions is conserved over the course of the reaction, and the ions maintain a $\sim$10 $\mu$m spacing in the Coulomb crystal, we can  conclude that there is no influence on the reactivity of the main reactions with acetylene. As a result of this contaminant reaction, data had to be normalized to the initial number of \ce{N2H+} minus the amount of protonated acetone present at each time point.  

While no reaction products were observed in the reactions between \ce{C6H5+} and either \ce{C2H2} or \ce{H2}, upper limits on the reaction rate constants for these reactions were determined by assuming \ce{C6H7+} products were observed using the following pseudo-first-order rate equations:
\begin{equation}
    \frac{dN_{77}}{dt} = -k_{calc}*C_{X}*N_{77}
\end{equation}
\begin{equation}
    \frac{dN_{prod}}{dt} = k_{calc}*C_{X}*N_{77}
\end{equation}
where N$_{prod}$ corresponds to the number of potential product ions from the reaction of \ce{C6H5+} with \ce{C2H2} or \ce{H2}, N$_{77}$ corresponds to the number of \ce{C6H5+} ions, C$_{X}$ is the concentration ( $\frac{molecules}{cm^{3}}$) of \ce{C2H2} or \ce{H2} present, and k$_{calc}$ corresponds to the reaction rate constant for the reaction between \ce{C6H5+} and \ce{C2H2} or \ce{H2}, depending on which reaction rate constant is being determined. This system of equations can be solved for k$_{calc}$ giving the following equation:
\begin{equation}
    k_{calc} = \frac{1}{t*C_{X}}*\ln{\frac{N_{77}(0)}{N_{77}(0)-N_{prod}(t)}}
\end{equation}
where t is the reaction time in seconds and N$_{77}$(0) is the initial amount of \ce{C6H5+} ($\sim90$ ions for the reaction with \ce{C2H2} and $\sim45$ ions for the reaction with \ce{H2}). Using both the pressures and reaction times mentioned previously along with an arbitrary signal of five product ions for N$_{prod}$(t), which is well above the noise floor, yields upper limits of $\sim4\times10^{-12}$ $\frac{cm^{3}}{s}$ and $\sim1\times10^{-12}$ $\frac{cm^{3}}{s}$ for the reaction with \ce{C2H2} and \ce{H2}, respectively.
It should be noted here that Bayard-Alpert ionization gauges have reduced accuracy below pressures of $10^{-8}$ Torr~\cite{singleton2001}; therefore, the reaction rate constants obtained here have some possible systematic uncertainty. The uncertainties provided represent statistical uncertainties in the measurements and do not reflect the accuracy of the ionization gauge.

\subsubsection*{Quantum chemical calculations}
Possible potential energy surfaces for the series of reactions studied here were calculated at the unrestricted MP2/aug-cc-pVTZ level using the Gaussian16 software package~\cite{Gaussian16}. Each surface begins with the reactants at infinite separation and progresses from one stationary point to the next until the final products are obtained. Stationary points were identified through the use of optimization and frequency calculations. Transition states, in particular, were identified by the presence of a vibrational mode with an imaginary frequency. Once a transition state had been found, the structure was altered along the direction of the imaginary frequency vibrational mode and a subsequent stationary point was determined. Only barrierless pathways were considered because of the low temperature environment of the experiment. The energy for each stationary point was zero-point energy corrected. The potential energy surfaces obtained from these calculations are provided in Supplementary Figures 1 and 2. In addition to these surfaces, molecular orbitals for the phenylium structure were calculated and shown in Supplementary Figure 4. The three $\pi$ bonding orbitals, outlined in the red boxes, indicate the presence of $4n+2$ H{\"u}ckel aromaticity. 

\bmhead{Supplementary information}
Supplementary information is provided in the Si.file. 

\bmhead{Acknowledgements}
The authors thank L.-S. Wang and M.-A. Martin-Drummel for helpful discussions during the preparation of this manuscript. This work was supported by the National Science Foundation (PHY-2317149, CHE-1900294) and the Air Force Office of Scientific Research (FA9550-20-1-0323).

\section*{Declarations}
\subsection*{Funding}
This work was supported by the Air Force Office of Scientific Research (FA9550-20-1-0323) and National Science Foundation (PHY-2317149, CHE-1900294).
\subsection*{Conflict of interest}
We declare no conflicts of interest

\subsection*{Data availability}
All data are available in the main text or in the supplementary information.
\subsection*{Author contribution}
Data collection and analysis were carried out by G.~S.~K. and C.~Z.-M. All authors contributed to interpreting the results and writing the manuscript.

\newpage
\section*{Supplementary Material}

\newpage
\renewcommand{\tablename}{Supplementary~Table}
\renewcommand{\figurename}{Supplementary~Fig.}
\setcounter{figure}{0}
\begin{table} 
	\centering
	
	\caption{\textbf{Reaction rate constants for each observed reaction.} Included are our measured values, as well as values from literature and Langevin theory. All units are in $\frac{cm^{3}}{s}$.The rate constants are calculated from fitting the rate equations to all data, with errors corresponding to a 90\% confidence interval.}
	\label{tab:rateconstants} 

	\begin{tabular}{lccr} 
		\\
		\hline
		Reaction & Observed & Literature &  Langevin\\
		\hline
        \ce{N2H+} + \ce{C2H2} &  $2.3 \pm 0.2\times10^{-9}$ & $1.4\times10^{-9}$ ~\cite{milligan2002PTRsm} & $1.18\times10^{-9}$\\
		\ce{C2H3+} + \ce{C2H2} & $9 \pm 1\times10^{-10}$ & $2.5\times10^{-10}$ ~\cite{kim1977carbocationssm} & $1.20\times10^{-9}$\\
		\ce{C4H3+} + \ce{C2H2} & $2.1 \pm 0.3\times10^{-10}$ & $\sim 2\times10^{-10}$ ~\cite{Derrick1970sm} & $1.05\times10^{-9}$\\
		\ce{C6H5+} + \ce{C2H2} & $<\sim4\times10^{-12}$ &  $\sim1\times10^{-11}$ ~\cite{knight1987SIFTsm} & $9.91\times10^{-10}$\\
        \ce{C6H5+} + \ce{H2} & $<\sim 1\times10^{-12}$ & $\sim3\times 10^{-11}$ ~\cite{ascenzi2003C6H5sm} &  $1.48\times10^{-9}$\\
		\hline
	\end{tabular}
\end{table}


\begin{figure}
    \centering
    \includegraphics[width=0.9\linewidth]{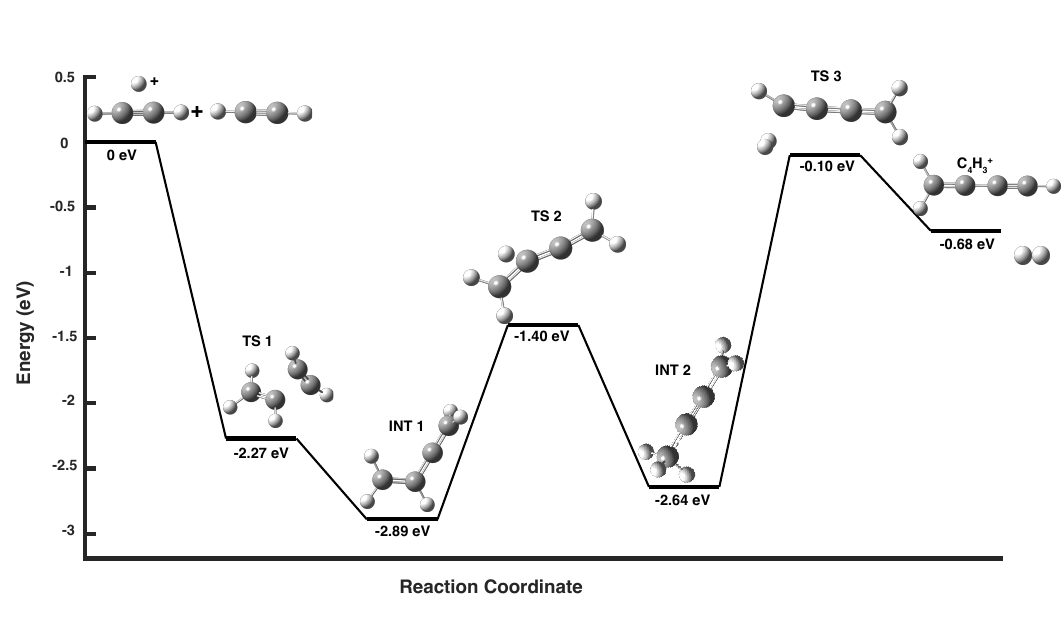}
    \caption{\textbf{Calculated potential energy surface to form \ce{C4H3+} from \ce{C2H3+} and \ce{C2H2}}.The surface begins with \ce{C2H3+} and \ce{C2H2} at infinite separation. The surface terminates upon formation of protonated diacetylene. Stationary points between these endpoints were identified using optimization and frequency calculations with transition states being identified by the presence of imaginary vibrational modes. All computations were completed at the MP2/aug-cc-pVTZ level of theory. The energy for each stationary point was zero-point energy corrected.}
    \label{fig:C4H3+PES}
\end{figure}

\begin{figure}
    \centering
    \includegraphics[width=0.9\linewidth]{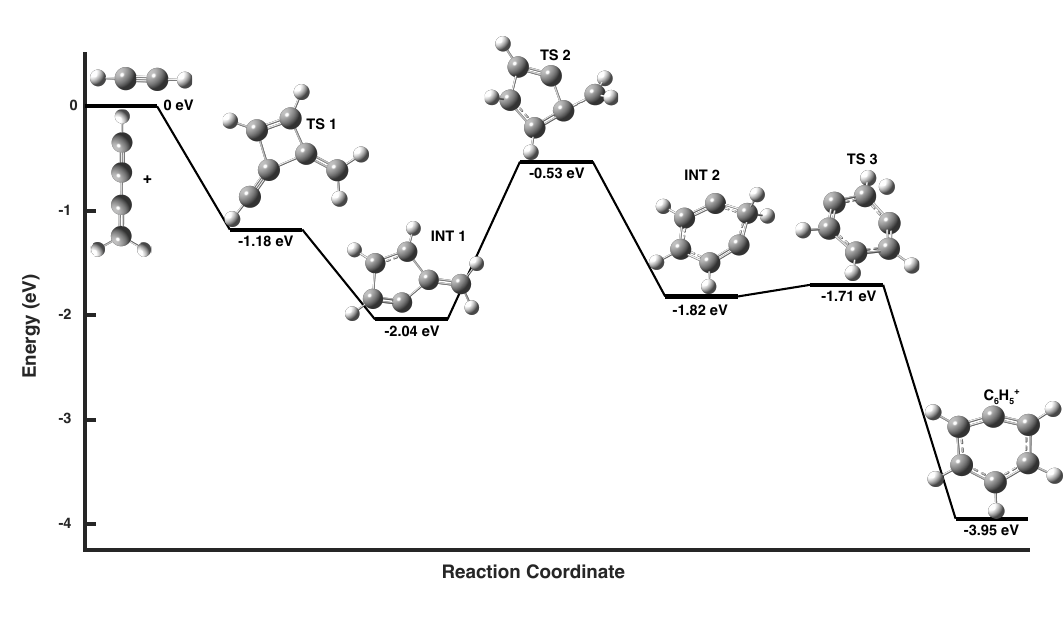}
    \caption{\textbf{Calculated potential energy surface to form \ce{C6H5+} from \ce{C4H3+} and \ce{C2H2}}.The surface begins with \ce{C4H3+} and \ce{C2H2} at infinite separation. The surface terminates upon formation of phenylium. Stationary points between these endpoints were identified using optimization and frequency calculations with transition states being identified by the presence of imaginary vibrational modes. All computations were completed at the MP2/aug-cc-pVTZ level of theory. The energy for each stationary point was zero-point energy corrected.}
    \label{fig:C6H5+PES}
\end{figure}

\begin{figure}
    \centering
    \includegraphics[width=0.9\linewidth]{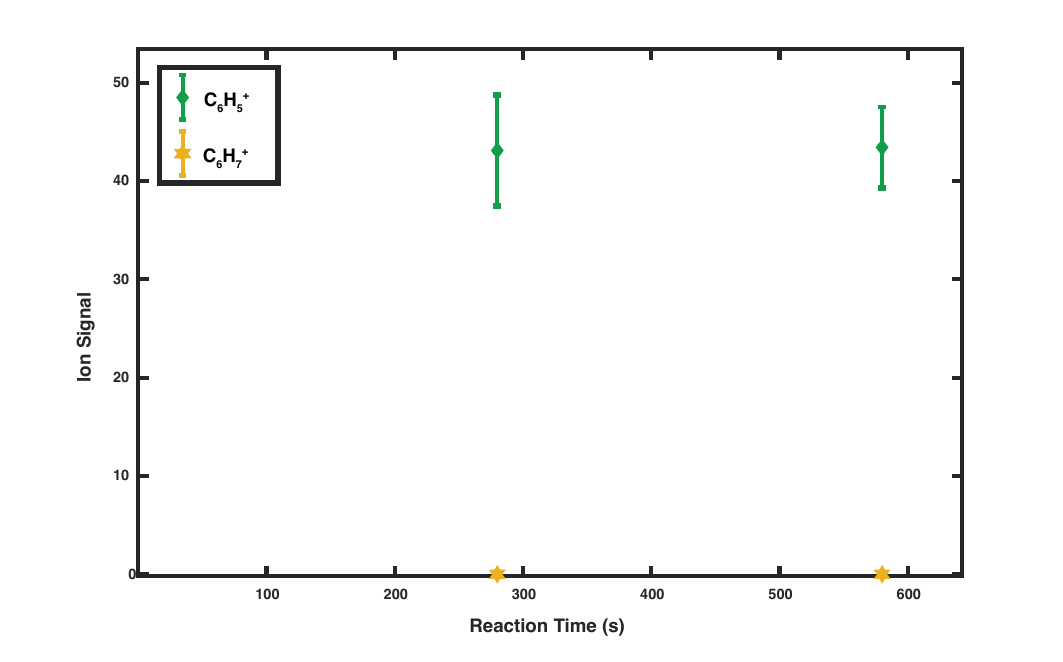}
    \caption{\textbf{Reaction results for \ce{C6H5+} + \ce{H2}.} After an initial period of 280 seconds to produce \ce{C6H5+}, the acetylene valve is closed and molecular hydrogen is introduced into the chamber for 300 seconds. Neither depletion of \ce{C6H5+} or growth of a new product, such as \ce{C6H7+}, are observed, indicating that no reaction occurs between \ce{C6H5+} and \ce{H2}.}
    \label{fig:H2RXN}
\end{figure}

\begin{figure}
    \centering
    \includegraphics[width=0.9\linewidth]{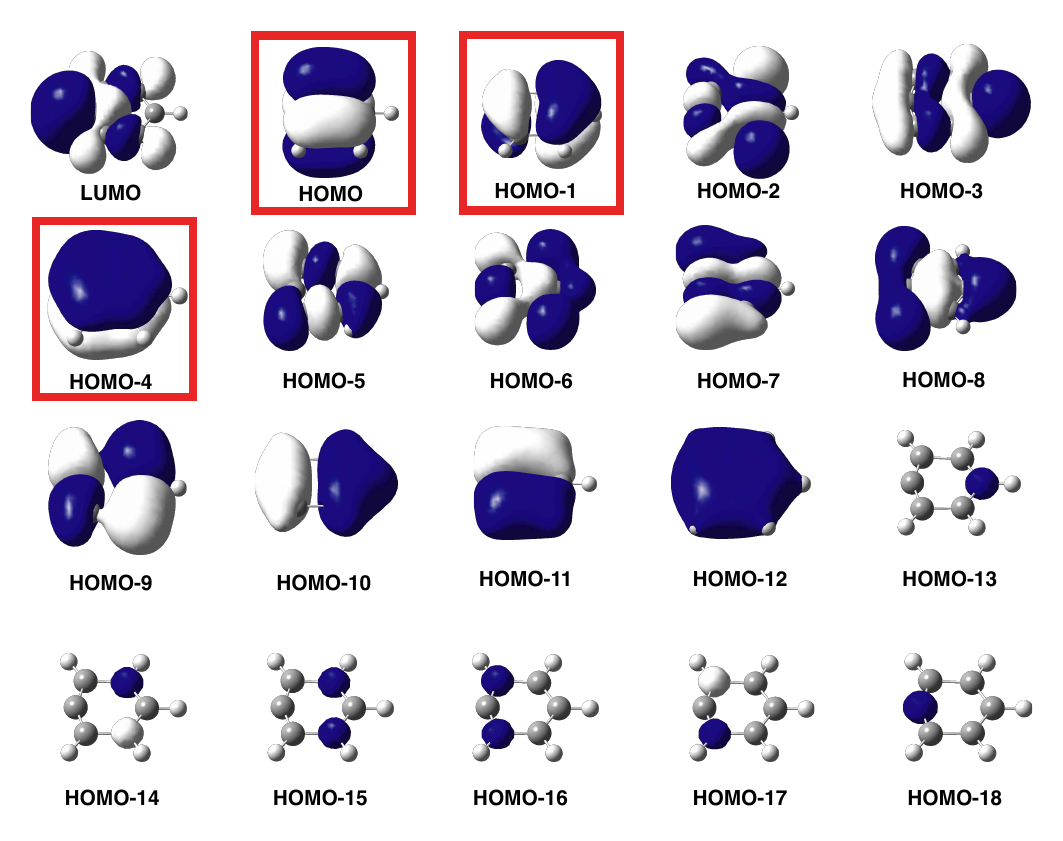}
    \caption{\textbf{Calculated molecular orbitals for phenylium}. The three delocalized $\pi$ orbitals that contribute to $4n+2$ H{\"u}ckel aromaticity are highlighted in red boxes.}
    \label{fig:MOs}
\end{figure}

\end{document}